\documentclass[11pt]{article}

\usepackage{amsmath}       
\usepackage[dvips]{graphicx}

\usepackage{multicol}

   \usepackage{sectsty}             
   \allsectionsfont{\sffamily}      

\newcommand{\mockalph}[1]{}

\usepackage[table]{xcolor}

\usepackage{natbib} 
\usepackage{url}

\graphicspath{{/home/andrew/same/rahema/paper/graphs/}}

\author{\textsf{Tina Ho}\thanks{Program in Public Health, %
    University of California, Irvine}\\
\textsf{Andrew Noymer}%
\thanks{Department of Population Health and Disease Prevention,
  University of California,
  Irvine}\textsf{\textsuperscript{\hspace{0.45em},}}%
\thanks{To whom correspondence should be addressed. 653 E Peltason
  Drive, Irvine CA 92697-3957, USA. \texttt{noymer@uci.edu}}%
\\ \texttt{noymer@uci.edu}}

\title{\vspace*{-3ex}%
\textsf{Summertime, and the livin' is easy:\\
    Winter and summer pseudoseasonal\\ life expectancy in the United
    States}%
}

\begin{document}
\maketitle

\thispagestyle{empty}


\vspace{-5ex}
\begin{abstract}
\noindent
In temperate climates, mortality is seasonal with a winter-dominant
pattern, due in part to pneumonia and influenza.  Cardiac causes,
which are the leading cause of death in the United States, are also
winter-seasonal although it is not clear why.  Interactions between
circulating respiratory viruses (f.e.\@, influenza) and cardiac
conditions have been suggested as a cause of winter-dominant mortality
patterns.  We propose and implement a way to estimate an upper bound
on mortality attributable to winter-dominant viruses like influenza.
We calculate `pseudo-seasonal' life expectancy, dividing the year into
two six-month spans, one encompassing winter the other summer.  During
the summer when the circulation of respiratory viruses is drastically
reduced, life expectancy is about one year longer.  We also quantify
the seasonal mortality difference in terms of seasonal ``equivalent
ages'' (defined herein) and proportional hazards.  We suggest that
even if viruses cause excess winter cardiac mortality, the
population-level mortality reduction of a perfect
influenza vaccine would be much more modest than is often recognized.\\
\vspace*{3ex}
\end{abstract}

\pagebreak

\noindent%
\hspace*{12em}\emph{Summertime,}\\
\hspace*{12em}\emph{And the livin' is easy}\\
\hspace*{12em}\emph{Fish are jumpin'}\\
\hspace*{12em}\emph{And the cotton is high}\\
\hspace*{12em}--- George Gershwin\\
\hspace*{12em}``Summertime''

{\begin{center}\noindent%
\rule[0.4ex]{\textwidth}{1.0pt}
\end{center}}

\vspace{2ex}

\section*{Introduction}

The primary goal of this paper is to forecast the best-case scenario
of life expectancy improvements that would accrue from the widespread
uptake of a perfect flu vaccine.  To accomplish this, we analyze life
expectancy in the United States from a seasonal perspective.  We
calculate two life expectancies per 12-month period (``pseudowinter''
and ``pseudosummer''), using methods described below.  The point is to
estimate life expectancy in the absence of respiratory viruses (most
notably, influenza), using pseudosummer as an approximation.
Pseudowinter, on the other hand, estimates life expectancy in the
presence of these viruses.  The difference between life expectancy in
pseudowinter and pseudosummer gives an upper bound on the potential
mortality impact of a perfect flu vaccine. The pseudoseasonal
approach also illuminates within-year mortality fluctuations.

Mortality in temperate climates is highly seasonal, with winter peaks
and summer troughs (\citealt{rosenberg-66}, \citealt{land-83},
\citealt{kalkstein-89}, \citealt{mackenbach-92}, \citealt{rau-book}).
Respiratory and cardiovascular causes, including stroke
(\citealt{sheth-99}), dominate the seasonal effects, with cancer being
negligibly cyclical (\citealt{crombie-95}).  Heat wave mortality peaks
are ephemeral interruptions of this overall pattern (f.e.\@
\citealt{basu-02}, \citealt{klinenberg-book}, \citealt{valleron-04},
\citealt{kaiser-07}, \citealt{toulemon-08}, \citealt{rocklov-11},
\citealt{robine-12}, \citealt{astrom-13}).  Heat wave-associated
deaths have a different composition by cause compared to summer
mortality (\citealt{huynen-01}, \citealt{basagana-11}), although
\cite{rey-07} report increases in most causes at older ages.  Heat
wave mortality interacts with air pollution (\citealt{rooney-98},
\citealt{bhaskaran-09}).  Both severity and duration of heat waves are
important for mortality (\citealt{anderson-09}).

Temperature is thought to play a role in mortality seasonality
(\citealt{braga-01, braga-02}, \citealt{curriero-02},
\citealt{mercer-03}).  However, temperature-associated deaths in a
literal sense (f.e.\@, hypothermia or heat stroke) are relatively
unimportant, with cold-related deaths slightly exceeding heat-related
deaths, at least in the United States (\citealt{berko-14}).
Nonetheless, the expansion over time of adequate winter heating in the
United States has been suggested as a possibly-overlooked factor in
the long-term decline of heart disease (\citealt{seretakis-97}).
Insufficient winter heating among the poor may not play a significant
role in mortality in Britain (\citealt{wilkinson-04}).
\cite{healy-03} demonstrate that the coefficient of seasonal variation
in mortality (CSVM) is correlated with mean winter temperature (warmer
temperature, higher CVSM) at the country level in Europe; see also
\cite{eurowinter-97} and \cite{diaz-05}.  \cite{analitis-08} also find
an association between cold weather and mortality in European cities,
and similarly note greater cold effect in warmer
climates. \cite{yang-12} and \cite{zhao-15} find similar results in
subtropical Asia. \cite{kysely-09} find increased cardiovascular
mortality in all ages above 25 during cold spells in the Czech
republic.  Mortality in nursing homes appears to be sensitive to both
hot and cold temperature extrema (\citealt{stafoggia-06},
\citealt{hajat-07}).

The role of temperature in mortality is an important topic in
historical demography, too large to survey completely here.  Much of
this work focuses on summer mortality, especially diarrhea among
infants and children (f.e.\@ \citealt{galloway-85},
\citealt{breschi-86a,breschi-86,breschi-86b}, \citealt{woods-89}).
There is a smaller body of work on winter peaks in infant mortality
before the twentieth century.  In particular, the hypothermia
hypothesis suggests that neonatal mortality increased in cold periods
(\citealt{dallazuanna-09, dallazuanna-11}; see also
\citealt{derosas-09,derosas-10} and \citealt{dallazuanna-10}).
Analyzing historical data from a cold-winter climate, \cite{astrom-16}
find that warmer spells are associated with lower mortality.
\cite{ekamper-09} find strong a strong social class influence on
temperature-mortality relationships in historical data from the
Netherlands, and review some of the literature on cold and mortality
in the past.

Cold temperature affects susceptibility to viruses in mice
\citep{foxman-15} and in human cells \emph{in vitro}
\citep{foxman-16}, although the evidence in humans is mixed
(\citealt{dowling-58}, \citealt{douglas-68}).  There may be
synergistic effects of temperature and humidity (\citealt{lowen-07},
\citealt{makinen-09}, \citealt{shaman-09}, \citealt{tebeest-13}).  In
the United States, mortality peaks coincide with the Christmas and New
Year's holidays, which occur during the northern hemisphere winter
(\citealt{phillips-04,phillips-10}).  However, Christmas effects on
cardiovascular mortality also occur in New Zealand, where the holiday
falls in the summertime (\citealt{knight-16}).  Hypovitaminosis D,
which is seasonal with winter peaks (\citealt{kasahara-13}), is also
thought to play a role in fatal diseases (\citealt{holick-07}).

The root causes of mortality seasonality remain poorly understood
(\citealt{dowell-01}, \citealt{cheng-05}).  There seems to be a nexus
between viral activity and adverse cardoiovascular events
(\citealt{bainton-78}, \citealt{kunst-93}, \citealt{madjid-04},
\citealt{huy-12}, \citealt{udell-13}).  However, the extent to which
respiratory virus transmission during the winter (\citealt{glezen-87})
causes increased mortality from other causes is debated
(\citealt{reichert-04}, \citealt{warren-gash-12},
\citealt{foster-13}).  The role of astronomical season (viz.\@,
through associated weather changes) in the cyclicality of infectious
disease is also debated (\citealt{fisman-12}, \citealt{treanor-16}),
with the school calendar (\citealt{grenfell-89}), and dynamic
resonance (\citealt{dushoff-04}) among alternate hypotheses.  The
plurality of explanations suggests to us that ``the mechanisms
underlying seasonality [of viral transmission] still remain
essentially unexplained'' (\citealt{yorke-79}, pp.~104--5).
Determining the \emph{causes} of seasonality of respiratory virus
transmission is beyond our scope.  Rather, we are concerned with
estimating the mortality \emph{consequences} of such seasonality.

\begin{figure}
\begin{center}
\includegraphics[width=.65\textwidth,keepaspectratio,clip=]{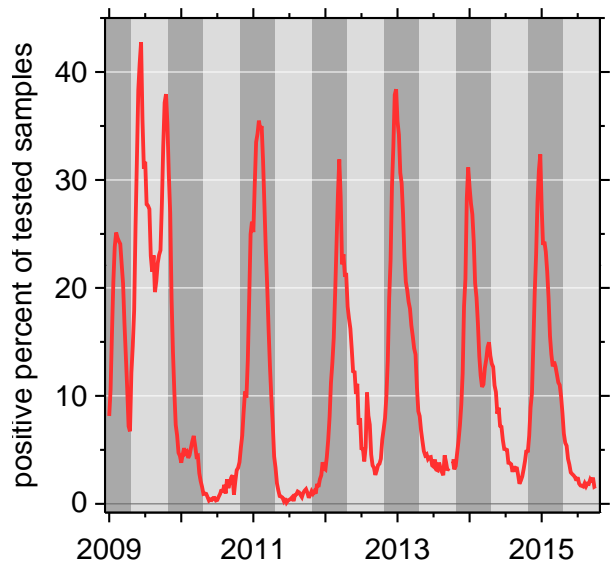}
\end{center}
\vspace{-4ex}
\caption{Percent of respiratory specimens testing positive for
  influenza virus, 2009--15.  Weekly data from \cite{fluview}. Pseudowinters
  are shaded dark. \label{PP}}
\end{figure}

\section*{Data and methods}

We present a simple and, to the best of our knowledge, novel, approach
to estimate the overall impact of winter-circulating viruses,
especially influenza, on mortality.  We divide the year into two
six-month ``pseudoseasons'', and calculate life expectancy for these
periods.  The seasonal binning approach using \emph{all-cause
  mortality} avoids potential classification pitfalls of
counterfactural approaches such as cause-deleted life tables or other
approaches which rely on cause of death reporting (f.e.,
\citealt{stewart-11}).  An influenza-deleted life expectancy is
calculated with flu mortality statistically removed
(\citealt{manton-86}), while our approach studies all-cause mortality,
but truly in the absence of the flu virus, i.e.\@, in the summertime.
The major strength of our approach is that our mortality estimates are
not hypothetical ``as if'' constructs, but reflect observed conditions
when no (or very little) flu virus circulates.  Among the problems
this avoids are classification errors regarding what is an influenza
death (cf.\@ \citealt{noymer-13}).

The noncirculation of flu viruses in the summertime is not absolute,
as figure~\ref{PP} shows.  This is a time series plot of respiratory
specimens (f.e.\@, nasal swabs) positive for any strain of the
influenza virus (as a percentage, so peaks are not reflective of more
samples during the winter).  Note that even in the peaks, most samples
test negative; there are many causes of upper respiratory illness
other than influenza virus.  Most peaks of influenza occur in the
shaded pseudowinters, but the 2009 swine-origin influenza pandemic is
a major exception.  During influenza pandemics, which involve
emergence of new strains, viral circulation in the summer is more
likely \citep{webster-92}.

From the mortality detail files of the National Center for Health
Statistics (NCHS \citeyear{mcd-data-alt-5}), we extracted monthly data
on every death in the United States, January 1959 to December
2014\footnote{Prior to 1959, digitized mortality data are not
  available for the United States that are simultaneously
  disaggreatable by age, sex, and month.}.  The data were then
aggregated by sex and 22 age groups
(0,1--4,5--9,...,95--99,$\geq$100), and binned into six-month
pseudoseasons.  Pseudowinter is November through April, and
pseudosummer is May through October; pseudoseasons do not nest into
calendar years.  In long-run averages, these six-month periods best
capture influenza virus circulation or lack thereof
(\citealt{thompson-09}).  The data begin with pseudosummer~1959 and
end with pseudosummer~2014 (56 pseudosummers).  There are~54
pseudowinters (1960--61 to~2013--14).  Data for January through April
1959 were discarded since using these data for pseudowinter 1959--60
would be biased due to the omission of November and December 1958.
Similarly, November and December 2014 were discarded.
We constructed denominators using age- and sex-specific calendar-year
exposure data from the \cite{hmd-16a-alt}.  We graduated these
person-years-at-risk data to months, adjusting for days per month and
leap years, and then re-aggregated to make pseudoseasonal exposures.  We
then calculated sex- and age-specific death rates for each
pseudoseason, from which we calculated sex-specific life tables in the
standard way (\citealt{keyfitz-70}, \citealt{preston-textbook}).

\section*{Results}

\begin{figure}
\begin{center}
\includegraphics[width=.85\textwidth,keepaspectratio,clip=]{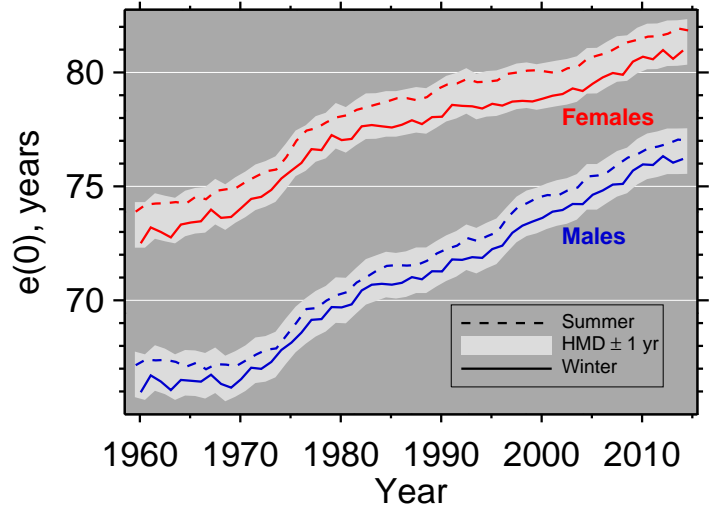}
\end{center}
\vspace{-4ex}
\caption{Life expectancy ($e(0)$) time series by sex and by
  pseudoseason.  The band enveloping the series is two years in
  height, centered on the calendar-year $e(0)$ estimates from
  \cite{hmd-16a-alt}; it is not an uncertainty interval. \label{fig1}}
\end{figure}

Figure~\ref{fig1} presents four $e(0)$ (life expectancy) time series:
pseudowinter (solid) and pseudosummer (dashed), for both males (lower
series, blue) and females (upper series, red).  The gray tubes
enveloping each sex are 2-year wide bands centered on calendar-year
life expectancy from the Human Mortality Database
(HMD)\nocite{hmd-16a-alt}; these are not uncertainty intervals. The
top of the gray band represents the calendar-year $e(0)$+1, so the
summer pseudoseasonal life expectancy is never greater than one year
above the neighboring calendar-year life expectancy.  Similarly, since
the bottom of the gray band is the calendar-year $e(0)-$1, it shows
that winter pseudoseasonal life expectancy is always within one year
of the neighboring calendar-year $e(0)$.  Using the HMD $e(0)$ data as
the center of the band also provides an external check of our life
expectancy calculations, since our pseudoseasonal data should fairly
neatly sandwich the calendar-year series.

\begin{figure}
\begin{center}
\includegraphics[width=.85\textwidth,keepaspectratio,clip=]{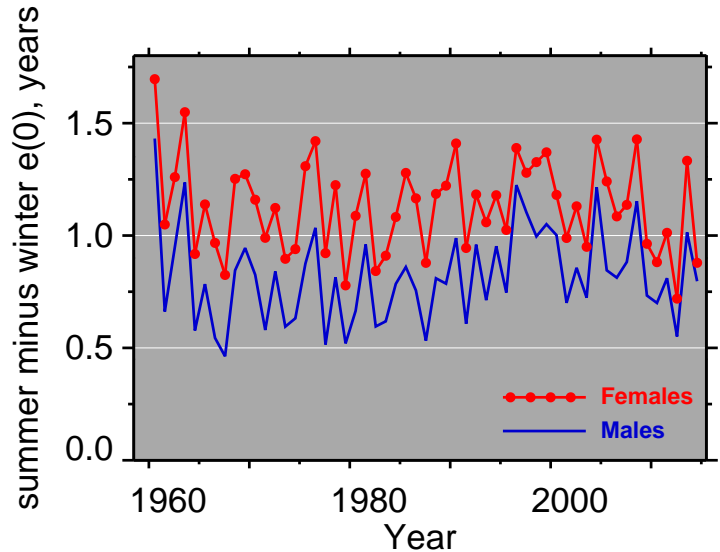}
\end{center}
\vspace{-4ex}
\caption{Summer advantage over winter in pseudoseasonal life
  expectancy, females (upper series) and males (lower
  series). \label{fig2}}
\end{figure}

Figure~\ref{fig2} shows the difference between $e(0)$ in summers and
their preceding winters (from the summer of 1960 minus the winter of
1959--60, to the summer of 2014 minus the winter of 2013--14).  There
are three important features.  First, no secular time trend is
evident.  Second, the data are strongly negatively autocorrelated:
declines are followed by increases, and vice versa.  Third, in
addition to higher life expectancy, women have a higher
summer$-$winter difference, 1.13$\pm$0.21 years, versus 0.82$\pm$0.21 years
for males (mean$\pm$SD).

\begin{figure}
\begin{center}
\includegraphics[width=\textwidth,keepaspectratio,clip=]{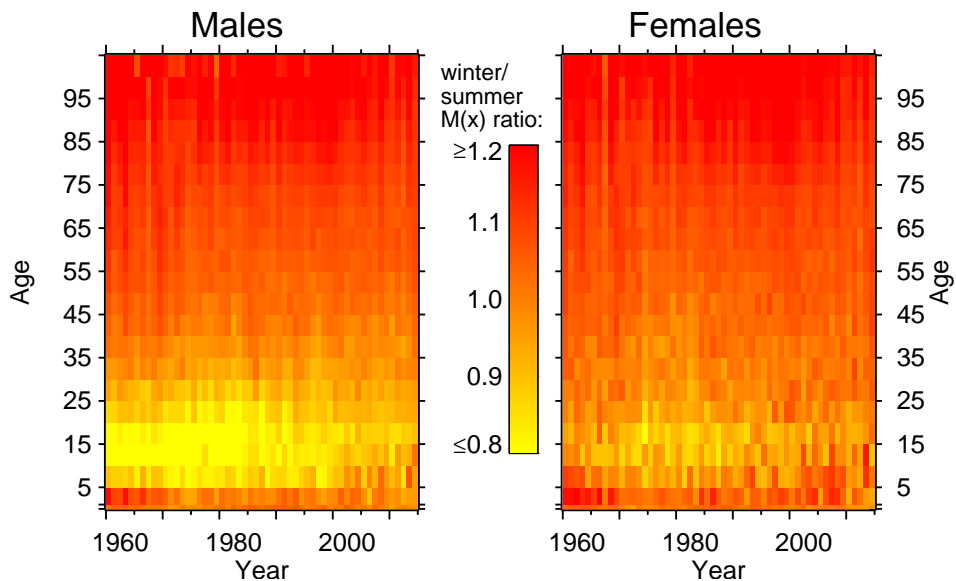}
\end{center}
\caption{Heatmap of winter:summer ratio, 1960--2014.  At younger ages
  (above childhood), summer has higher mortality, although this has
  decreased over time.  At older ages,  winters have higher mortality.
  \label{heatmap}}
\end{figure}

Figure~\ref{heatmap} is a heat map of the winter:summer ratio of the
mortality rate by age ($M_x$), over time.  Several features of
figure~\ref{heatmap} are especially relevant to seasonal differences.
First, summer advantage in mortality is an age-related phenomenon. At
younger ages (approximately~5--35), summers are more deadly.  The
summer excess is more pronounced for males, and is declining over
time.  It is particularly noticeable in the so-called accident bump
\citep{pampel-01}.  Indeed, summer mortality at younger ages is
associated with motor vehicle fatalities \citep{farmer-05} and
external causes generally \citep{feinstein-02}.  Winter overtakes
summer above age 45, where death rates are (much) higher in absolute
terms.

\begin{figure}
\begin{center}
\includegraphics[width=.85\textwidth,keepaspectratio,clip=]{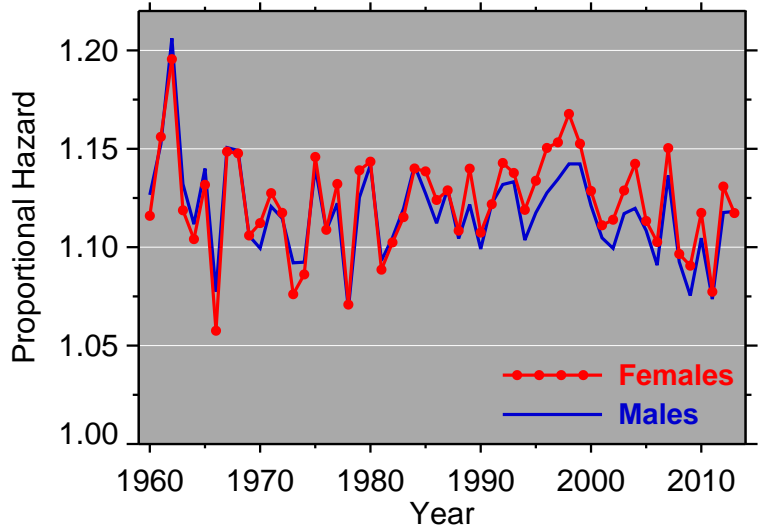}
\end{center}
\vspace{-4ex}
\caption{Proportional hazard analysis. The proportional hazard is
  winter death rates above age~45 as a multiple of summer
  death rates. 
  \label{PH}}
\end{figure}

Figure~\ref{PH} helps quantify the pseudoseasonal differences seen in
figure~\ref{heatmap}.  Here we present, on a year-by-year basis, the
proportional hazard ($P^Y$) of summer mortality for age$\geq45$,
separately by sex.  Thus, we model $\mathbf{W}=P^Y\mathbf{S}$ where
$\mathbf{W}$ is the $N$-element (agewise) vector of winter death rates
for a given year, $P^Y$ is the year-specific proportional hazard
(scalar), and $\mathbf{S}$ is the vector of summer death rates.  The
proportional hazard is estimated as:
\begin{equation}
P^Y=%
\exp\left(\left.\sum_{x=45}^{\omega}\left[\log\big(M^W_x\big)
-\log\big(M^S_x\big)\right]\middle/N\right. \right)
\label{estimation}
\end{equation}
where the superscripts ($W, S$) refer to winter and summer and $M_x$
is the age-specific death rate.  The proportional hazard is the same
as the winter to summer ratio of the geometric mean death rate (see
\citealt{schoen-70}).  Figure~\ref{PH} shows that most winters have a
mortality pattern that is between 110\%--115\% of the previous
summer'\kern-0.05em{}s mortality. The proportional hazard model is an
excellent fit, with all the year-specific $R^2>0.99$, which is not
especially surprising, since it is based on pairwise comparisons of
adjacent pseudoseasons.  There are no meaningful sex differences in
the proportional hazard.\footnote{This is in contrast to Denmark,
  where \cite{rau-03} find greater seasonal fluctuations for males.}
If we drop the $Y$ superscript and model a single proportional hazard
for the entire data set, then $\mathbf{W}$ and $\mathbf{S}$ become
year $\times$ age matrices, and $P$ is 1.119 for males and 1.124 for
females.  Naturally, when modeling the whole time span with a single
$P$, the goodness of fit declines, but it is still not poor:
$R^2=0.89$ for males and $R^2=0.88$ for females.

\section*{Discussion}

Taking only mortality into account, how much happier should an adult
be during the summer?  Death rates will be, typically, 10\% to 15\%
higher in the winter (figure~\ref{PH}).  However, by the time summer
arrives, up to half a year will have passed, and death rates will be
higher due to aging, even half a year's worth.  How does winter:summer
mortality difference compare to age-related changes?  We propose
calculating ``equivalent ages''\kern-0.1em, as follows.  In
table~\ref{gom-tab}, the $M_x$ columns give death rates by age, sex,
and pseudoseason.  The ``w.e.a.\@'' columns give the \emph{winter
  equivalent age}, or the age at which one would have to be in the
winter to experience the same (summer) death rate.  Similarly, the
``s.e.a.\@'' columns give the \emph{summer equivalent age}, or the
summer age that experiences the same (winter) death rate. The $M_x$,
w.e.a.\@ and s.e.a.\@ are calculated from a Gompertz mortality model
estimated by Poisson regression (cf.\@ \citealt{abdullatif}, p.~207),
the coefficients of which are given at the top of the table.
Symbolically:
\begin{eqnarray}
  M_x^S&=&\exp\big(\hat\alpha^S+\hat\beta^S\cdot{}x^S\big) \label{sum}\\ 
  M_x^W&=&\exp\big(\hat\alpha^W+\hat\beta^W\cdot{}x^W\big) \label{win}\\ 
  \text{w.e.a.\@}(x^S)&=&%
 \big(\hat\alpha^S-\hat\alpha^W+\hat\beta^S\cdot{}x^S)/\hat\beta^W  
       \label{wea}
\end{eqnarray}
where $S, W$ superscripts are for summer and winter, $x$ is age, and
$\hat\alpha, \hat\beta$, are estimated coefficients.  The solution for
$\text{w.e.a.\@}(x^S)$ in (\ref{wea}) comes from setting mortality
rates (i.e., [\ref{sum}] and [\ref{win}]) equal, and solving for $x^W$
in terms of $x^S$ and the estimated coefficients. Thus, if
$\text{w.e.a.\@}(x^S)$ is plugged into \ref{win} for $x^W$, it will
produce a death rate equivalent to the desired $M_x^S$. The same
formula holds, \emph{mutatis mutandis}, for $\text{s.e.a.\@}(x^W)$.
The (winter/summer) equivalent age is a function of the estimated
Gompertz coefficients for both pseudoseasons and of the age for which
an equivalency is being calculated.

Table~\ref{gom-tab} gives specific examples, using 2010 data. An 80
year old woman in the winter experiences death rates of an 81 year old
woman in the summer.  Death rates are higher in the winter, so the
equivalent age in the summer is older. The difference is one year of
age.  On the other hand, a 70 year old man living in the summer has
death rates equivalent to a 69.1 year old man in the winter.  Summer
mortality is more lenient and therefore it's as if he is a younger
man, compared to winter.  The absolute value of the difference between
biological age and w.e.a.\@ or s.e.a.\@ becomes larger as biological
age increases, since $M_x$ increases exponentially with age.

There is a micro-macro disconnect here: for populations, mortality is
clearly lower in the summer, holding age constant.  From the point of
view of an individual, holding age constant is meaningless; one cannot
go from winter to summer without aging approximately half a
year. Thus, the winter-into-summer mortality changes experienced by an
individual are less than the \emph{ceteris paribus} analysis
represented by the proportional hazards (and the summer-into-winter
changes, more).

\begin{table}

\definecolor{lightgray}{gray}{0.8}
\rowcolors{7}{}{lightgray}

\begin{center}
\hspace*{-3em}%
\begin{tabular}{r||ccc|ccc|ccc|cc}
\multicolumn{2}{c}{}& \multicolumn{4}{c}{\textbf{Women}} &&&  \multicolumn{4}{c}{\textbf{Men}} \\
\cline{3-6}\cline{9-12}
\multicolumn{2}{c}{}& \multicolumn{2}{c}{SUMMER}&\multicolumn{2}{c}{WINTER} &&&  
    \multicolumn{2}{c}{SUMMER}&\multicolumn{2}{c}{WINTER} \\
   && $\hat\alpha=$& $-$10.94& $\hat\alpha=$&$-$10.95&&&     
      $\hat\alpha=$& $-$9.80 & $\hat\alpha=$&$-$9.85  \\
($x$)   && $\hat\beta=$& .0975& $\hat\beta=$&.0989 &&&    
      $\hat\beta=$& .0869& $\hat\beta=$&.0888       \\ 
\cline{3-6}\cline{9-12}
Age && $M_x$ & w.e.a.\@& $M_x$ & s.e.a.\@&&& $M_x$ & w.e.a.\@& $M_x$ & s.e.a.\@\\
\hline\hline
50 &&   231.1 & 49.40 &   245.1 & 50.60 &&&   428.6 & 49.53 &   446.8 & 50.48 \\
60 &&   612.6 & 59.27 &   658.7 & 60.74 &&&  1,021.9 & 59.32 &  1,085.5 & 60.70 \\
70 &&  1,623.7 & 69.13 &  1,770.0 & 70.89 &&&  2,436.0 & 69.11 &  2,637.0 & 70.91 \\
80 &&  4,303.9 & 78.99 &  4,756.4 & 81.03 &&&  5,807.4 & 78.89 &  6,406.2 & 81.13 \\
90 && 11,408.1 & 88.85 & 12,781.3 & 91.17 &&& 13,844.4 & 88.68 & 15,563.1 & 91.35 \\
\hline
\end{tabular}
\end{center}
\caption{Equivalent age analysis for 2010, as explained in the main text. $M_x$:
  modeled death rate per 100,000; w.e.a.\@ is ``winter equivalent age'', or the
  age at which an individual would experience the same death rate,
  living in the winter, and s.e.a.\@ is the same, \emph{mutatis mutandis},
  for living in the summer.  Coefficients from a Poisson regression; 
  the prediction equation is: $M_x=\exp(\alpha+\beta{}x)$.
  \label{gom-tab}}
\end{table}

In terms of life expectancy, the effect of the winter increase in
mortality is similarly modest: on average, just over one year of life
expectancy for women and just under one year for men.  If we could
wave a magic wand, eradicating influenza, respiratory syntictal virus,
and other pathogens which circulate in the winter, and, what is more,
making the winter pattern of cardiac mortality look like the summer
pattern (regardless of the reason for its seasonality), this would be
equivalent to about seven years' worth (in terms of time) of recent
mortality progress (i.e.\@, based on the slopes of figure~\ref{fig1}).

The reason for this modest difference is easy to see, at least in
retrospect.  If we could eradicate influenza, then it would be like
living in the summer.  To put it another way, people would experience
their ``summer equivalent age'', as in the example above.  Although
reducing death rates by about 12\% seems like a great thing, it only
makes a small difference in equivalent age, and, therefore, has a
modest effect on $e(0)$.  Mindel Sheps's (\citeyear{sheps-58})
observation that changes in death rates usually are smaller when
viewed through the lens of concomitant changes in survival rates, is
highly relevant.  The appreciable pseudoseasonal difference in $M_x$
results in a rather modest difference in $e(0)$ because life
expectancy is the integral of the life table $\ell_x$, or survivor,
column, not the $M_x$ column.  The Gompertzian relationship that
holds above age 45, where by far the majority of deaths occur,
guarantees that age-associated increases in mortality would swamp the
hypothetical change generated by eradication of influenza.

The novelty of our approach lies not in the idea that elimination of a
seemingly-major cause (in this case, approximated by pseudosummer)
will have a small impact on $e(0)$.  This is well understood; for
example, \cite{keyfitz-amd} (pp.~62--72) considers it in relation to
the Shannon entropy, $H$, of the life table $\ell_x$ column.  Because
of competing risks of, say, heart disease, even eradicating cancer
does not cause huge changes in $e(0)$ (\emph{ibid.}), so it is clear
that removing influenza deaths also won't have a big effect.
\citeauthor{vaupel-86}'s study (\citeyear{vaupel-86}) of the relation
between $e(0)$ and $M_x$ is also relevant.  The greater mortality
seasonality of women as measured by life expectancy differences
(figure~\ref{fig2}) than as measured by the winter:summer proportional
hazard (figure~\ref{PH}), is consistent with this; the effect of a
constant multiple of $M_x$ affects $e(0)$ differently at different
levels of $M_x$.  What the present study shows, however, is that the
\emph{total} mortality impact of influenza (viz.\@, including knock-on
effects of flu on heart disease) is not very large in the grand scheme
of things.

Another approach to estimating the role of viruses in all-cause
mortality would be to use direct measures of viral circulation (as in
figure~\ref{PP}) instead of summer and winter as instrumental
indicators.  This would have the disadvantage of not being applicable
to historical data (viral surveillance like that shown in
figure~\ref{PP} begins in the late 1990s).  On the other hand, an
advantage is that it could be applied to the tropics, where influenza
circulation is more haphazard (see f.e.\@ \citealt{aungkulanon-15}),
and therefore the colinearity of flu season and ``winter'' is neither
an appropriate identification strategy nor a lurking problem.  Using
nominal influenza mortality as an instrument, instead of the seasons,
is another possibility, but is not without its problems
\citep{noymer-13}.

This study has a number of strengths and weaknesses.  The principal
strength is that it uses all-cause mortality and so automatically
includes any influenza-related deaths that would ordinarily be missed.
One limitation is that we can only observe summers that follow
winters, and vice versa.  The more lenient mortality of the summer
leads to the accumulation of frail individuals who then become more
likely to die in the winter.  Similarly, the more severe mortality of
the winter leaves a more robust residual population, less likely to
die in the summer; here we use ``robust'' and ``frail'' in the usual
demographic sense (\citealt{keyfitz-79}, \citealt{vaupel-79},
\citealt{vaupel-yashin-85}, \citealt{manton-86}).  Thus, diminution of
winter mortality from the invention of a perfect influenza vaccine
could precipitate small \emph{increases} in summer mortality, as a
result of perturbing the frail/robust cycle.  \cite{goldstein-12}
(p.~833) likewise speculate that these effects are limited in
magnitude.  The negative autocorrelation seen in figures~\ref{fig2}
and~\ref{PH} may well be driven by similar effects.  It is also
possible that influenza and other respiratory pathogens are
under-ascertained in the summertime, and thus that the roots of
seasonal mortality are misunderstood, although this seems unlikely
given figure~\ref{PP} and other work \citep{hayward-14}.

\section*{Conclusion}

It is reasonably well established that at least part of the reduction
in cardiovascular mortality during the summer is due to the absence of
influenza virus (\citealt{warren-gash-09,warren-gash-11}).  The
connection between influenza vaccine use and reduction of wintertime
heart mortality is less clear, but has been studied
(\citealt{seo-14}).  Influenza vaccine use is correlated with general
health-seeking behavior, and so confounding is a problem in a direct
empirical approaches to this question.  In this study, we took an
indirect approach, using whole-population data, and summers as a sort
of natural experiment.

The question of mortality in a world with much more effective flu
vaccines is not purely of theoretical importance.  One of the chief
reasons influenza vaccine is not optimally effective is the constant
evolution of the virus (\citealt{treanor-04}).  This leads directly to
two related obstacles to good population-level immunity: the need to
reformulate the flu vaccine each year, with not all years having equal
vaccine efficacy (\citealt{keitel-97}), and the need for people to be
revaccinated each year.  Progress is being made toward a vaccine that
solves both of these problems \citep{pica-13}.  A universal flu
vaccine (as such shots are called) is a clearly-expressed public
health desideratum (\citealt{fineberg-14}), but remains on the drawing
board.  Realistic expectations for mortality changes should be part of
the policy analysis in this area.

In conclusion, the effect of influenza on life expectancy in the
United States is less than 1.25 years for women and less than 1.0 year
for men.  This estimate is based on summer-winter differences and so
implicitly includes the knock-on effect of influenza on other causes,
most notably heart disease. This must be regarded as an upper bound on
the gains to life expectancy from a universal flu vaccine, which
could --- theoretically --- eradicate influenza, but not other
winter-circulating respiratory pathogens.  The morality impact of such
a vaccine would be neither negligible nor enormous.

\subsection*{Acknowledgments}

The idea for this paper is an offshoot of a stimulating conversation
with Viggo Andreasen.  For helpful suggestions, we thank Bob Schoen and
Monica He King, as well as seminar audiences at Ohio State, Universit\'e de
Montr\'eal, and both the Health Policy Research Institute and the
Institute for Mathematical Behavioral Sciences at UC, Irvine.  Carter
Butts suggested the title.  Rahema Haseeb provided research assistance.

\begin{multicols}{2}
{ \scriptsize
\bibliography{refs}

\begin{thebibliography}{110}
\newcommand{\enquote}[1]{``#1''}
\providecommand{\natexlab}[1]{#1}
\providecommand{\url}[1]{\texttt{#1}}
\providecommand{\urlprefix}{URL }

\bibitem[\protect\citeauthoryear{\AA{}str\"om et~al.}{2016}]{astrom-16}
\AA{}str\"om, Daniel~Oudin, S\"oren Edvinsson, Daniel Hondula, Joacim
  Rockl\"ov, and Barbara Schumann. 2016.
\newblock \enquote{On the association between weather variability and total and
  cause-specific mortality before and during industrialization in {S}weden.}
\newblock \emph{Demographic Research} 35(33):991--1010.

\bibitem[\protect\citeauthoryear{\AA{}str\"om et~al.}{2013}]{astrom-13}
\AA{}str\"om, Daniel~Oudin, Bertil Forsberg, S\"oren Edvinsson, and Joacim
  Rockl\"ov. 2013.
\newblock \enquote{Acute fatal effects of short-lasting extreme temperatures in
  {S}tockholm, {S}weden: Evidence across a century of change.}
\newblock \emph{Epidemiology} 24(6):820--829.

\bibitem[\protect\citeauthoryear{Abdullatif and Noymer}{2016}]{abdullatif}
Abdullatif, Viytta~N. and Andrew Noymer. 2016.
\newblock \enquote{\emph{Clostridium difficile} infection: An emerging cause of
  death in the twenty-first century.}
\newblock \emph{Biodemography and Social Biology} 62(2):198--207.

\bibitem[\protect\citeauthoryear{Analitis et~al.}{2008}]{analitis-08}
Analitis, A., K.~Katsouyanni, A.~Biggeri, M.~Baccini, B.~Forsberg, L.~Bisanti,
  U.~Kirchmayer, F.~Ballester, E.~Cadum, P.~G. Goodman, A.~Hojs, J.~Sunyer,
  P.~Tiittanen, and P.~Michelozzi. 2008.
\newblock \enquote{Effects of cold weather on mortality: Results from 15
  {E}uropean cities within the {PHEWE} project.}
\newblock \emph{American Journal of Epidemiology} 168(12):1397--1408.

\bibitem[\protect\citeauthoryear{Anderson and Bell}{2009}]{anderson-09}
Anderson, Brooke~G. and Michelle~L. Bell. 2009.
\newblock \enquote{Weather-related mortality: How heat, cold, and heat waves
  affect mortality in the {U}nited {S}tates.}
\newblock \emph{Epidemiology} 20(2):205--213.

\bibitem[\protect\citeauthoryear{Aungkulanon et~al.}{2015}]{aungkulanon-15}
Aungkulanon, Suchunya, Po-Yung Cheng, Khanitta Kusreesakul, Kanitta
  Bundhamcharoen, Malinee Chittaganpitch, McCarron Margaret, and Sonja Olsen.
  2015.
\newblock \enquote{Influenza-associated mortality in {T}hailand, 2006--2011.}
\newblock \emph{Influenza and Other Respiratory Viruses} 9(6):298--304.

\bibitem[\protect\citeauthoryear{Bainton et~al.}{1978}]{bainton-78}
Bainton, David, Glynne~R Jones, and David Hole. 1978.
\newblock \enquote{Influenza and ischaemic heart disease -- A possible trigger
  for acute myocardial infarction?}
\newblock \emph{International Journal of Epidemiology} 7(3):231--239.

\bibitem[\protect\citeauthoryear{Basaga\~{n}a et~al.}{2011}]{basagana-11}
Basaga\~{n}a, Xavier, Claudio Sartini, Jose Barrera-G\'omez, Payam Dadvand,
  Jordi Cunillera, Bart Ostro, Jordi Sunyer, and Mercedes Medina-Ramon. 2011.
\newblock \enquote{Heat waves and cause-specific mortality at all ages.}
\newblock \emph{Epidemiology} 22(6):765--772.

\bibitem[\protect\citeauthoryear{Basu and Samet}{2002}]{basu-02}
Basu, Rupa and Jonathan~M. Samet. 2002.
\newblock \enquote{Relation between elevated ambient temperature and mortality:
  A review of the epidemiologic evidence.}
\newblock \emph{Epidemiologic Reviews} 24(2):190--202.

\bibitem[\protect\citeauthoryear{Berko et~al.}{2014}]{berko-14}
Berko, Jeffrey, Deborah~D. Ingram, Shubhayu Saha, and Jennifer~D. Parker. 2014.
\newblock \enquote{Deaths attributed to heat, cold, and other weather events in
  the {U}nited {S}tates, 2006--2010.}
\newblock \emph{National Health Statistics Reports} 76.

\bibitem[\protect\citeauthoryear{Bhaskaran et~al.}{2009}]{bhaskaran-09}
Bhaskaran, K., S.~Hajat, A.~Haines, E.~Herrett, P.~Wilkinson, and L.~Smeeth.
  2009.
\newblock \enquote{Effects of ambient temperature on the incidence of
  myocardial infarction.}
\newblock \emph{Heart} 95(21):1760--1769.

\bibitem[\protect\citeauthoryear{Braga et~al.}{2002}]{braga-02}
Braga, Alf\'esio L.~F., Antonella Zanobetti, and Joel Schwartz. 2002.
\newblock \enquote{The effect of weather on respiratory and cardiovascular
  deaths in 12 {U.S.} cities.}
\newblock \emph{Environmental Health Perspectives} 110(9):859--863.

\bibitem[\protect\citeauthoryear{Braga et~al.}{2001}]{braga-01}
Braga, Alf\'esio Lu\'{\i}s~Ferreira, Antonella Zanobetti, and Joel Schwartz.
  2001.
\newblock \enquote{The time course of weather-related deaths.}
\newblock \emph{Epidemiology} 12(6):662--667.

\bibitem[\protect\citeauthoryear{Breschi and
  Livi-Bacci}{1986{\natexlab{\emph{a}}}}]{breschi-86a}
Breschi, Marco and Massimo Livi-Bacci. 1986{\natexlab{\emph{a}}}.
\newblock \enquote{{\mockalph{A}}Saison et climat comme contraintes de la
  survie des enfants: L'exp\'erience italienne au XIXe si\`ecle.}
\newblock \emph{Population} 41(1):9--35.

\bibitem[\protect\citeauthoryear{Breschi and
  Livi-Bacci}{1986{\natexlab{\emph{b}}}}]{breschi-86}
---{}---{}---. 1986{\natexlab{\emph{b}}}.
\newblock \enquote{{\mockalph{B}}Stagione di nascita e clima come determinanti
  della mortalita' infantile negli stati sardi di terraferma.}
\newblock \emph{Genus} 42(1/2):87--101.

\bibitem[\protect\citeauthoryear{Breschi and
  Livi-Bacci}{1986{\natexlab{\emph{c}}}}]{breschi-86b}
---{}---{}---. 1986{\natexlab{\emph{c}}}.
\newblock \enquote{{\mockalph{C}}Effect du climat sur la mortalit\'e infantile:
  R\'esultats pour la Savoie, le Pi\'emont et la Ligurie en 1828--1837.}
\newblock \emph{Population} 41(6):1072--1074.

\bibitem[\protect\citeauthoryear{CDC}{2015}]{fluview}
CDC. 2015.
\newblock \emph{Weekly {U.S.} Influenza Surveillance Report}.
\newblock \url{http://www.cdc.gov/flu/weekly/}.
\newblock Accessed 9 June 2016.

\bibitem[\protect\citeauthoryear{Cheng}{2005}]{cheng-05}
Cheng, Tsung~O. 2005.
\newblock \enquote{Mechanism of seasonal variation in acute myocardial
  infarction.}
\newblock \emph{International Journal of Cardiology} 100(1):163--164.

\bibitem[\protect\citeauthoryear{Crombie et~al.}{1995}]{crombie-95}
Crombie, D~L, D~M Fleming, K~W Cross, and R~J Lancashire. 1995.
\newblock \enquote{Concurrence of monthly variations of mortality related to
  underlying cause in {E}urope.}
\newblock \emph{Journal of Epidemiology and Community Health} 49(4):373--378.

\bibitem[\protect\citeauthoryear{Curriero et~al.}{2002}]{curriero-02}
Curriero, Frank~C., Karlyn~S. Heiner, Jonathan~M. Samet, Scott~L. Zeger, Lisa
  Strug, and Jonathan~A. Patz. 2002.
\newblock \enquote{Temperature and mortality in 11 cities of the eastern
  {U}nited {S}tates.}
\newblock \emph{American Journal of Epidemiology} 155(1):80--87.

\bibitem[\protect\citeauthoryear{Dalla-Zuanna and
  Rosina}{2009}]{dallazuanna-09}
Dalla-Zuanna, Gianpiero and Alessandro Rosina. 2009.
\newblock \enquote{The fatal season: An analysis of extremely high winter
  neonatal mortality.}
\newblock \emph{Transylvanian Review} 18(1):245--276.

\bibitem[\protect\citeauthoryear{Dalla-Zuanna and
  Rosina}{2010}]{dallazuanna-10}
---{}---{}---. 2010.
\newblock \enquote{A note on: The joint effect of maternal malnutrition and
  cold weather on neonatal mortality in -century {V}enice: An assessment of the
  hypothermia hypothesis, \emph{Population Studies} 63(3):233--251 by Renzo
  Derosas.}
\newblock \emph{Population Studies} 64(2):193--195.

\bibitem[\protect\citeauthoryear{Dalla-Zuanna and
  Rosina}{2011}]{dallazuanna-11}
---{}---{}---. 2011.
\newblock \enquote{An analysis of extremely high nineteenth-century winter
  neonatal mortality in a local context of northeastern {I}taly.}
\newblock \emph{European Journal of Population} 27(1):33--55.

\bibitem[\protect\citeauthoryear{Derosas}{2009}]{derosas-09}
Derosas, Renzo. 2009.
\newblock \enquote{The joint effect of maternal malnutrition and cold weather
  on neonatal mortality in nineteenth-century {V}enice: An assessment of the
  hypothermia hypothesis.}
\newblock \emph{Population Studies} 63(3):233--251.

\bibitem[\protect\citeauthoryear{Derosas}{2010}]{derosas-10}
---{}---{}---. 2010.
\newblock \enquote{Reply to the note by {Dalla-Zuanna} and {R}osina.}
\newblock \emph{Population Studies} 64(2):197--198.

\bibitem[\protect\citeauthoryear{D\'\i{}az et~al.}{2005}]{diaz-05}
D\'\i{}az, Julio, Ricardo Garc\'\i{}a, C\'esar L\'opez, Cristina Linares,
  Aurelio Tob\'\i{}as, and Luis Prieto. 2005.
\newblock \enquote{Mortality impact of extreme winter temperatures.}
\newblock \emph{International Journal of Biometeorology} 49(3):179--183.

\bibitem[\protect\citeauthoryear{Douglas et~al.}{1968}]{douglas-68}
Douglas, R.~Gordon, Jr., Keith~M. Lindgren, and Robert~B. Couch. 1968.
\newblock \enquote{Exposure to cold environment and rhinovirus common cold.}
\newblock \emph{New England Journal of Medicine} 279(14):742--747.

\bibitem[\protect\citeauthoryear{Dowell}{2001}]{dowell-01}
Dowell, Scott~F. 2001.
\newblock \enquote{Seasonal variation in host susceptibility and cycles of
  certain infectious diseases.}
\newblock \emph{Emerging Infectious Diseases} 7(3):369--374.

\bibitem[\protect\citeauthoryear{Dowling et~al.}{1958}]{dowling-58}
Dowling, Harry~F., George~Gee Jackson, Irwin~G. Spiesman, and Tohru Inouye.
  1958.
\newblock \enquote{Transmission of the common cold to volunteers under
  controlled conditions. III. The effect of chilling of the subjects upon
  susceptibility.}
\newblock \emph{American Journal of Hygiene} 68(1):59--65.

\bibitem[\protect\citeauthoryear{Du\-shoff et~al.}{2004}]{dushoff-04}
Du\-shoff, Jonathan, Joshua~B. Plotkin, Simon~A. Levin, and David J.~D. Earn.
  2004.
\newblock \enquote{Dynamical resonance can account for seasonality of influenza
  epidemics.}
\newblock \emph{Proceedings of the National Academy of Sciences of the United
  States of America} 101(48):16,915--16,916.

\bibitem[\protect\citeauthoryear{Ekamper et~al.}{2009}]{ekamper-09}
Ekamper, Peter, Frans van Poppel, Coen van Duin, and Joop Garssen. 2009.
\newblock \enquote{150 Years of temperature-related excess mortality in the
  {N}etherlands.}
\newblock \emph{Demographic Research} 21(14):385--426.

\bibitem[\protect\citeauthoryear{Farmer and Williams}{2005}]{farmer-05}
Farmer, C.~M. and A.~F. Williams. 2005.
\newblock \enquote{Temporal factors in motor vehicle crash deaths.}
\newblock \emph{Injury Prevention} 11(1):18--23.

\bibitem[\protect\citeauthoryear{Feinstein}{2002}]{feinstein-02}
Feinstein, Craig~A. 2002.
\newblock \enquote{Seasonality of deaths in the {U.S.} by age and cause.}
\newblock \emph{Demographic Research} 6(17):469--486.

\bibitem[\protect\citeauthoryear{Fineberg}{2014}]{fineberg-14}
Fineberg, Harvey~V. 2014.
\newblock \enquote{Pandemic preparedness and response: Lessons from the {H1N1}
  influenza of 2009.}
\newblock \emph{New England Journal of Medicine} 370(14):1335--1342.

\bibitem[\protect\citeauthoryear{Fisman}{2012}]{fisman-12}
Fisman, D. 2012.
\newblock \enquote{Seasonality of viral infections: Mechanisms and unknowns.}
\newblock \emph{Clinical Microbiology and Infection} 18(10):946--954.

\bibitem[\protect\citeauthoryear{Foster et~al.}{2013}]{foster-13}
Foster, E.~D., J.~E. Cavanaugh, W.~G. Haynes, M.~Yang, A.~K. Gerke, F.~Tang,
  and P.~M. Polgreen. 2013.
\newblock \enquote{Acute myocardial infarctions, strokes and influenza:
  Seasonal and pandemic effects.}
\newblock \emph{Epidemiology and Infection} 141(4):735--744.

\bibitem[\protect\citeauthoryear{Foxman et~al.}{2015}]{foxman-15}
Foxman, Ellen~F., James~A. Storer, Megan~E. Fitzgerald, Bethany~R. Wasik, Lin
  Hou, Hongyu Zhao, Paul~E. Turner, Anna~Marie Pyle, and Akiko Iwasaki. 2015.
\newblock \enquote{Temperature-dependent innate defense against the common cold
  virus limits viral replication at warm temperature in mouse airway cells.}
\newblock \emph{Proceedings of the National Academy of Sciences of the United
  States of America} 112(3):827--832.

\bibitem[\protect\citeauthoryear{Foxman et~al.}{2016}]{foxman-16}
Foxman, Ellen~F., James~A. Storer, Kiran Vanaja, Andre Levchenko, and Akiko
  Iwasaki. 2016.
\newblock \enquote{Two interferon-independent double-stranded {RNA}-induced
  host defense strategies suppress the common cold virus at warm temperature.}
\newblock \emph{Proceedings of the National Academy of Sciences of the United
  States of America} 113(30):8496--8501.

\bibitem[\protect\citeauthoryear{Galloway}{1985}]{galloway-85}
Galloway, P.~R. 1985.
\newblock \enquote{Annual variations in deaths by age, deaths by cause, prices,
  and weather in {L}ondon 1670 to 1830.}
\newblock \emph{Population Studies} 39(3):487--505.

\bibitem[\protect\citeauthoryear{Glezen et~al.}{1987}]{glezen-87}
Glezen, W.~Paul, Michael Decker, Sheldon~W. Joseph, and Raymond G.~Mercready
  Jr. 1987.
\newblock \enquote{Acute respiratory disease associated with influenza
  epidemics in {H}ouston, 1981--1983.}
\newblock \emph{Journal of Infectious Diseases} 155(6):1119--1126.

\bibitem[\protect\citeauthoryear{Goldstein et~al.}{2012}]{goldstein-12}
Goldstein, Edward, Cecile Viboud, Vivek Charu, and Marc Lipsitch. 2012.
\newblock \enquote{Improving the estimation of influenza-related mortality over
  a seasonal baseline.}
\newblock \emph{Epidemiology} 23(6):829--838.

\bibitem[\protect\citeauthoryear{Grenfell and Anderson}{1989}]{grenfell-89}
Grenfell, B.~T. and R.~M. Anderson. 1989.
\newblock \enquote{Pertussis in {E}ngland and {W}ales: An investigation of
  transmission dynamics and control by mass vaccination.}
\newblock \emph{Proceedings of the Royal Society of London, Series B:
  Biological Sciences} 236(1284):pp. 213--252.

\bibitem[\protect\citeauthoryear{Hajat et~al.}{2007}]{hajat-07}
Hajat, S., R.~S. Kovats, and K.~Lachowycz. 2007.
\newblock \enquote{Heat-related and cold-related deaths in {E}ngland and
  {W}ales: Who is at risk?}
\newblock \emph{Occupational and Environmental Medicine} 64(2):93--100.

\bibitem[\protect\citeauthoryear{Hayward et~al.}{2014}]{hayward-14}
Hayward, Andrew~C., Ellen~B. Fragaszy, Alison Bermingham, Lili Wang, Andrew
  Copas, W.~John Edmunds, Neil Ferguson, Nilu Goonetilleke, Gabrielle Harvey,
  Jana Kovar, Megan S.~C. Lim, Andrew McMichael, Elizabeth R.~C. Millett,
  Jonathan~S. Nguyen-Van-Tam, Irwin Nazareth, Richard Pebody, Faiza Tabassum,
  John~M Watson, Fatima~B. Wurie, Anne~M. Johnson, and Maria Zambon. 2014.
\newblock \enquote{Comparative community burden and severity of seasonal and
  pandemic influenza: Results of the {F}lu {W}atch cohort study.}
\newblock \emph{Lancet Respiratory Medicine} 2(6):445--454.

\bibitem[\protect\citeauthoryear{Healy}{2003}]{healy-03}
Healy, J.~D. 2003.
\newblock \enquote{Excess winter mortality in {E}urope: A cross country
  analysis identifying key risk factors.}
\newblock \emph{Journal of Epidemiology and Community Health} 57(10):784--789.

\bibitem[\protect\citeauthoryear{Holick}{2007}]{holick-07}
Holick, Michael~F. 2007.
\newblock \enquote{Vitamin {D} deficiency.}
\newblock \emph{New England Journal of Medicine} 357(3):266--281.

\bibitem[\protect\citeauthoryear{Human Mortality Database}{2016}]{hmd-16a-alt}
Human Mortality Database. 2016.
\newblock \url{http://www.mortality.org/}.
\newblock {Accessed 15 August 2016}.

\bibitem[\protect\citeauthoryear{Huy et~al.}{2012}]{huy-12}
Huy, Christina, Dorothee Kuhn, Sven Schneider, and Iris Z\"ollner. 2012.
\newblock \enquote{Seasonal waves of influenza and cause-specific mortality in
  {G}ermany.}
\newblock \emph{Central European Journal of Medicine} 7(4):450--456.

\bibitem[\protect\citeauthoryear{Huynen et~al.}{2001}]{huynen-01}
Huynen, Maud M. T.~E., Pim Martens, Dieneke Schram, Matty~P. Weijenberg, and
  Anton~E. Kunst. 2001.
\newblock \enquote{The impact of heat waves and cold spells on mortality rates
  in the {D}utch population.}
\newblock \emph{Environmental Health Perspectives} 109(5):463--470.

\bibitem[\protect\citeauthoryear{Kaiser et~al.}{2007}]{kaiser-07}
Kaiser, Reinhard, Alain {Le Tertre}, Joel Schwartz, Carol~A. Gotway,
  W.~Randolph Daley, and Carol~H. Rubin. 2007.
\newblock \enquote{The effect of the 1995 heat wave in {C}hicago on all-cause
  and cause-specific mortality.}
\newblock \emph{American Journal of Public Health} 97(S1):S158--S162.

\bibitem[\protect\citeauthoryear{Kalkstein and Davis}{1989}]{kalkstein-89}
Kalkstein, Laurence~S. and Robert~E. Davis. 1989.
\newblock \enquote{Weather and human mortality: An evaluation of demographic
  and interregional responses in the {U}nited {S}tates.}
\newblock \emph{Annals of the Association of American Geographers}
  79(1):44--64.

\bibitem[\protect\citeauthoryear{Kasahara et~al.}{2013}]{kasahara-13}
Kasahara, Amy~K., Ravinder~J. Singh, and Andrew Noymer. 2013.
\newblock \enquote{Vitamin {D} ({25OHD}) serum seasonality in the {U}nited
  {S}tates.}
\newblock \emph{PLoS One} 8(6):e65,785.

\bibitem[\protect\citeauthoryear{Keatinge et~al.}{1997}]{eurowinter-97}
Keatinge, W.~R., G.~C. Donaldson, K.~Bucher, G.~Jendritsky, E.~Cordioli,
  M.~Martinelli, L.~Dardanoni, K.~Katsouyanni, A.~E. Kunst, J.~P. Mackenbach,
  C.~McDonald, S.~Nayha, and I.~Vuori. 1997.
\newblock \enquote{Cold exposure and winter mortality from ischaemic heart
  disease, cerebrovascular disease, respiratory disease, and all causes in warm
  and cold regions of {E}urope.}
\newblock \emph{Lancet} 349(9062):1341--1346.

\bibitem[\protect\citeauthoryear{Keitel et~al.}{1997}]{keitel-97}
Keitel, Wendy~A., Thomas~R. Cate, Robert~B. Couch, Linda~L. Huggins, and
  Kenneth~R. Hesst. 1997.
\newblock \enquote{Efficacy of repeated annual immunization with inactivated
  influenza virus vaccines over a five year period.}
\newblock \emph{Vaccine} 15(10):1114--1122.

\bibitem[\protect\citeauthoryear{Keyfitz and Littman}{1979}]{keyfitz-79}
Keyfitz, N. and G.~Littman. 1979.
\newblock \enquote{Mortality in a heterogeneous population.}
\newblock \emph{Population Studies} 33(2):333--342.

\bibitem[\protect\citeauthoryear{Keyfitz}{1970}]{keyfitz-70}
Keyfitz, Nathan. 1970.
\newblock \enquote{Finding probabilities from observed rates, or how to make a
  life table.}
\newblock \emph{American Statistician} 24(1):28--33.

\bibitem[\protect\citeauthoryear{Keyfitz}{1985}]{keyfitz-amd}
---{}---{}---. 1985.
\newblock \emph{Applied mathematical demography}.
\newblock Springer, New York, second ed.

\bibitem[\protect\citeauthoryear{Klinenberg}{2002}]{klinenberg-book}
Klinenberg, Eric. 2002.
\newblock \emph{Heat wave: A social autopsy of disaster in {C}hicago}.
\newblock University of Chicago Press.

\bibitem[\protect\citeauthoryear{Knight et~al.}{2016}]{knight-16}
Knight, Josh, Chris Schilling, Adrian Barnett, Rod Jackson, and Phillip Clarke.
  2016.
\newblock \enquote{Revisiting the `{C}hristmas holiday effect' in the southern
  hemisphere.}
\newblock \emph{Journal of the American Heart Association} 5(12):{e005,098}.

\bibitem[\protect\citeauthoryear{Kunst et~al.}{1993}]{kunst-93}
Kunst, Anton~E., Casper W.~N. Looman, and Johan~P. Mackenbach. 1993.
\newblock \enquote{Outdoor air temperature and mortality in the {N}etherlands:
  A time-series analysis.}
\newblock \emph{American Journal of Epidemiology} 137(3):331--341.

\bibitem[\protect\citeauthoryear{Kysely et~al.}{2009}]{kysely-09}
Kysely, Jan, Lucie Pokorna, Jan Kyncl, and Bohumir Kriz. 2009.
\newblock \enquote{Excess cardiovascular mortality associated with cold spells
  in the {C}zech {R}epublic.}
\newblock \emph{BMC Public Health} 9(1):19.

\bibitem[\protect\citeauthoryear{Land and Cantor}{1983}]{land-83}
Land, Kenneth~C. and David Cantor. 1983.
\newblock \enquote{{ARIMA} models of seasonal variation in U.S.\@ birth and
  death rates.}
\newblock \emph{Demography} 20(4):541--568.

\bibitem[\protect\citeauthoryear{Lowen et~al.}{2007}]{lowen-07}
Lowen, Anice~C, Samira Mubareka, John Steel, and Peter Palese. 2007.
\newblock \enquote{Influenza virus transmission is dependent on relative
  humidity and temperature.}
\newblock \emph{PLoS Pathogens} 3(10):e151.

\bibitem[\protect\citeauthoryear{Mackenbach et~al.}{1992}]{mackenbach-92}
Mackenbach, J.~P., A.~E. Kunst, and C.~W. Looman. 1992.
\newblock \enquote{Seasonal variation in mortality in The {N}etherlands.}
\newblock \emph{Journal of Epidemiology and Community Health} 46(3):261--265.

\bibitem[\protect\citeauthoryear{Madjid et~al.}{2004}]{madjid-04}
Madjid, Mohammad, Ibrahim Aboshady, Imran Awan, Silvio Litovsky, and S.~Ward
  Casscells. 2004.
\newblock \enquote{Influenza and cardiovascular disease: Is there a causal
  relationship?}
\newblock \emph{Texas Heart Institute Journal} 31(1):4--13.

\bibitem[\protect\citeauthoryear{M\"{a}kinen et~al.}{2009}]{makinen-09}
M\"{a}kinen, Tiina~M., Raija Juvonen, Jari Jokelainen, Terttu~H. Harju, Ari
  Peitso, Aini Bloigu, Sylvi Silvennoinen-Kassinen, Maija Leinonen, and Juhani
  Hassi. 2009.
\newblock \enquote{Cold temperature and low humidity are associated with
  increased occurrence of respiratory tract infections.}
\newblock \emph{Respiratory Medicine} 103(3):456--462.

\bibitem[\protect\citeauthoryear{Manton et~al.}{1986}]{manton-86}
Manton, Kenneth~G., Eric Stallard, and James~W. Vaupel. 1986.
\newblock \enquote{Alternative models for the heterogeneity of mortality risks
  among the aged.}
\newblock \emph{Journal of the American Statistical Association}
  81(395):635--644.

\bibitem[\protect\citeauthoryear{Mercer}{2003}]{mercer-03}
Mercer, James~B. 2003.
\newblock \enquote{Cold---An underrated risk factor for health.}
\newblock \emph{Environmental Research} 92(1):8--13.

\bibitem[\protect\citeauthoryear{{National Center for Health
  Statistics}}{2015}]{mcd-data-alt-5}
{National Center for Health Statistics}. 2015.
\newblock \emph{Mortality multiple cause-of-death data files,
  \emph{\url{http://www.cdc.gov/nchs/nvss/mortality_public_use_data.htm}}}.
\newblock National Center for Health Statistics.
\newblock Accessed 12 December 2015.

\bibitem[\protect\citeauthoryear{Noymer and Nguyen}{2013}]{noymer-13}
Noymer, Andrew and Ann~M. Nguyen. 2013.
\newblock \enquote{Influenza as a proportion of pneumonia mortality: {U}nited
  {S}tates, 1959--2009.}
\newblock \emph{Biodemography and Social Biology} 59(2):178--190.

\bibitem[\protect\citeauthoryear{Pampel}{2001}]{pampel-01}
Pampel, Fred~C. 2001.
\newblock \enquote{Gender equality and the sex differential in mortality from
  accidents in high income nations.}
\newblock \emph{Population Research and Policy Review} 20(5):397--421.

\bibitem[\protect\citeauthoryear{Phillips et~al.}{2010}]{phillips-10}
Phillips, David, Gwendolyn~E. Barker, and Kimberly~M. Brewer. 2010.
\newblock \enquote{{C}hristmas and {N}ew {Y}ear as risk factors for death.}
\newblock \emph{Social Science \& Medicine} 71(8):1463--1471.

\bibitem[\protect\citeauthoryear{Phillips et~al.}{2004}]{phillips-04}
Phillips, David~P., Jason~R. Jarvinen, Ian~S. Abramson, and Rosalie~R.
  Phillips. 2004.
\newblock \enquote{Cardiac mortality is higher around {C}hristmas and {N}ew
  {Y}ear's than at any other time: The holidays as a risk factor for death.}
\newblock \emph{Circulation} 110(25):3781--3788.

\bibitem[\protect\citeauthoryear{Pica and Palese}{2013}]{pica-13}
Pica, Natalie and Peter Palese. 2013.
\newblock \enquote{Toward a universal influenza virus vaccine: Prospects and
  challenges.}
\newblock \emph{Annual Review of Medicine} 64(1):189--202.

\bibitem[\protect\citeauthoryear{Preston et~al.}{2001}]{preston-textbook}
Preston, Samuel~H., Patrick Heuveline, and Michel Guillot. 2001.
\newblock \emph{Demography: Measuring and modeling population processes}.
\newblock Blackwell, Oxford.

\bibitem[\protect\citeauthoryear{Rau}{2006}]{rau-book}
Rau, Roland. 2006.
\newblock \emph{Seasonality in human mortality: A demographic approach}.
\newblock No.~3 in Demographic Research Monographs, Springer, Berlin.

\bibitem[\protect\citeauthoryear{Rau and Doblhammer}{2003}]{rau-03}
Rau, Roland and Gabriele Doblhammer. 2003.
\newblock \enquote{Seasonal mortality in {D}enmark: The role of sex and age.}
\newblock \emph{Demographic Research} 9(9):197--222.

\bibitem[\protect\citeauthoryear{Reichert et~al.}{2004}]{reichert-04}
Reichert, Thomas~A., Lone Simonsen, Ashutosh Sharma, Scott~A. Pardo, David~S.
  Fedson, and Mark~A. Miller. 2004.
\newblock \enquote{Influenza and the winter increase in mortality in the
  {U}nited {S}tates, 1959--1999.}
\newblock \emph{American Journal of Epidemiology} 160(5):492--502.

\bibitem[\protect\citeauthoryear{Rey et~al.}{2007}]{rey-07}
Rey, Gr\'egoire, Anne Fouillet, \'Eric Jougla, and Denis H\'emon. 2007.
\newblock \enquote{Heat waves, ordinary temperature fluctuations and mortality
  in {F}rance since 1971.}
\newblock \emph{Population-E} 62(3):457--485.

\bibitem[\protect\citeauthoryear{Robine et~al.}{2012}]{robine-12}
Robine, {J.-M.}, {J.-P.} Michel, and F.~R. Herrmann. 2012.
\newblock \enquote{Excess male mortality and age-specific mortality
  trajectories under different mortality conditions: A lesson from the heat
  wave of summer 2003.}
\newblock \emph{Mechanisms of Ageing and Development} 133(6):378--386.

\bibitem[\protect\citeauthoryear{Rockl\"ov et~al.}{2011}]{rocklov-11}
Rockl\"ov, Joacim, Kristie Ebi, and Bertil Forsberg. 2011.
\newblock \enquote{Mortality related to temperature and persistent extreme
  temperatures: A study of cause-specific and age-stratified mortality.}
\newblock \emph{Occupational and Environmental Medicine} 68(7):531--536.

\bibitem[\protect\citeauthoryear{Rooney et~al.}{1998}]{rooney-98}
Rooney, Cleone, Anthony~J. McMichael, R.~Sari Kovats, and Michel~P. Coleman.
  1998.
\newblock \enquote{Excess mortality in {E}ngland and {W}ales, and in {G}reater
  {L}ondon, during the 1995 heatwave.}
\newblock \emph{Journal of Epidemiology and Community Health} 52(8):482--486.

\bibitem[\protect\citeauthoryear{Rosenberg}{1966}]{rosenberg-66}
Rosenberg, Harry~M. 1966.
\newblock \enquote{Recent developments in seasonally adjusting vital
  statistics.}
\newblock \emph{Demography} 3(2):305--318.

\bibitem[\protect\citeauthoryear{Schoen}{1970}]{schoen-70}
Schoen, Robert. 1970.
\newblock \enquote{The geometric mean of the age-specific death rates as a
  summary index of mortality.}
\newblock \emph{Demography} 7(3):317--324.

\bibitem[\protect\citeauthoryear{Seo et~al.}{2014}]{seo-14}
Seo, Yu~Bin, Won~Suk Choi, Ji~Hyeon Baek, Jacob Lee, Joon~Young Song, Jin~Soo
  Lee, Hee~Jin Cheong, and Woo~Joo Kim. 2014.
\newblock \enquote{Effectiveness of the influenza vaccine at preventing
  hospitalization due to acute exacerbation of cardiopulmonary disease in
  {K}orea from 2011 to 2012.}
\newblock \emph{Human Vaccines \& Immunotherapeutics} 10(2):423--427.

\bibitem[\protect\citeauthoryear{Seretakis et~al.}{1997}]{seretakis-97}
Seretakis, Dimitrios, Pagona Lagiou, Loren Lipworth, Lisa~B. Signorello,
  Kenneth~J. Rothman, and Dimitrios Trichopoulos. 1997.
\newblock \enquote{Changing seasonality of mortality from coronary heart
  disease.}
\newblock \emph{Journal of the American Medical Association}
  278(12):1012--1014.

\bibitem[\protect\citeauthoryear{Shaman and Kohn}{2009}]{shaman-09}
Shaman, Jeffrey and Melvin Kohn. 2009.
\newblock \enquote{Absolute humidity modulates influenza survival,
  transmission, and seasonality.}
\newblock \emph{Proceedings of the National Academy of Sciences of the United
  States of America} 106(9):3243--3248.

\bibitem[\protect\citeauthoryear{Sheps}{1958}]{sheps-58}
Sheps, Mindel~C. 1958.
\newblock \enquote{Shall we count the living or the dead?}
\newblock \emph{New England Journal of Medicine} 259(25):1210--1214.

\bibitem[\protect\citeauthoryear{Sheth et~al.}{1999}]{sheth-99}
Sheth, Tej, Cyril Nair, James Muller, and Salim Yusuf. 1999.
\newblock \enquote{Increased winter mortality from acute myocardial infarction
  and stroke: The effect of age.}
\newblock \emph{Journal of the American College of Cardiology}
  33(7):1916--1919.

\bibitem[\protect\citeauthoryear{Stafoggia et~al.}{2006}]{stafoggia-06}
Stafoggia, Massimo, Francesco Forastiere, Daniele Agostini, Annibale Biggeri,
  Luigi Bisanti, Ennio Cadum, Nicola Caranci, Francesca de'Donato, Sara
  De~Lisio, Moreno De~Maria, Paola Michelozzi, Rossella Miglio, Paolo Pandolfi,
  Sally Picciotto, Magda Rognoni, Antonio Russo, Corrado Scarnato, and Carlo~A.
  Perucci. 2006.
\newblock \enquote{Vulnerability to heat-related mortality: A multicity,
  population-based, case-crossover analysis.}
\newblock \emph{Epidemiology} 17(3):315--323.

\bibitem[\protect\citeauthoryear{Stewart}{2011}]{stewart-11}
Stewart, Quincy~Thomas. 2011.
\newblock \enquote{The cause-deleted index: Estimating cause of death
  contributions to mortality.}
\newblock \emph{Mathematical Population Studies} 18(4):234--257.

\bibitem[\protect\citeauthoryear{{te Beest} et~al.}{2013}]{tebeest-13}
{te Beest}, Dennis~E., Michiel {van Boven}, Mari\"ette Hooiveld, Carline {van
  den Dool}, and Jacco Wallinga. 2013.
\newblock \enquote{Driving factors of influenza transmission in the
  {N}etherlands.}
\newblock \emph{American Journal of Epidemiology} 178(9):1469--1477.

\bibitem[\protect\citeauthoryear{Thompson et~al.}{2009}]{thompson-09}
Thompson, William~W., Eric Weintraub, Praveen Dhankhar, Po-Yung Cheng, Lynnette
  Brammer, Martin~I. Meltzer, Joseph~S. Bresee, and David~K. Shay. 2009.
\newblock \enquote{Estimates of {US} influenza-associated deaths made using
  four different methods.}
\newblock \emph{Influenza and Other Respiratory Viruses} 3(1):37--49.

\bibitem[\protect\citeauthoryear{Toulemon and Barbieri}{2008}]{toulemon-08}
Toulemon, Laurent and Magali Barbieri. 2008.
\newblock \enquote{The mortality impact of the August 2003 heat wave in
  {F}rance: Investigating the `harvesting' effect and other long-term
  consequences.}
\newblock \emph{Population Studies} 62(1):39--53.

\bibitem[\protect\citeauthoryear{Treanor}{2004}]{treanor-04}
Treanor, John. 2004.
\newblock \enquote{Influenza vaccine: Outmaneuvering antigenic shift and
  drift.}
\newblock \emph{New England Journal of Medicine} 350(3):218--220.

\bibitem[\protect\citeauthoryear{Treanor}{2016}]{treanor-16}
Treanor, John~J. 2016.
\newblock \enquote{Influenza vaccination.}
\newblock \emph{New England Journal of Medicine} 375(13):1261--1268.

\bibitem[\protect\citeauthoryear{Udell et~al.}{2013}]{udell-13}
Udell, Jacob~A., Rami Zawi, Deepak~L. Bhatt, Maryam Keshtkar-Jahromi, Fiona
  Gaughran, Arintaya Phrommintikul, Andrzej Ciszewski, Hossein Vakili,
  Elaine~B. Hoffman, Michael~E. Farkouh, and Christopher~P. Cannon. 2013.
\newblock \enquote{Association between influenza vaccination and cardiovascular
  outcomes in high-risk patients: A meta-analysis.}
\newblock \emph{Journal of the American Medical Association}
  310(16):1711--1720.

\bibitem[\protect\citeauthoryear{Valleron and Boumendil}{2004}]{valleron-04}
Valleron, Alain-Jacques and Ariane Boumendil. 2004.
\newblock \enquote{\'Epid\'emiologie et canicules: Analyses de la vague de
  chaleur 2003 en {F}rance.}
\newblock \emph{Comptes Rendus Biologies} 327(12):1125--1141.

\bibitem[\protect\citeauthoryear{Vaupel}{1986}]{vaupel-86}
Vaupel, J.~W. 1986.
\newblock \enquote{How change in age-specific mortality affects life
  expectancy.}
\newblock \emph{Population Studies} 40(1):147--157.

\bibitem[\protect\citeauthoryear{Vaupel et~al.}{1979}]{vaupel-79}
Vaupel, James~W., Kenneth~G. Manton, and Eric Stallard. 1979.
\newblock \enquote{The impact of heterogeneity in individual frailty on the
  dynamics of mortality.}
\newblock \emph{Demography} 16(3):439--454.

\bibitem[\protect\citeauthoryear{Vaupel and Yashin}{1985}]{vaupel-yashin-85}
Vaupel, James~W. and Anatoli~I. Yashin. 1985.
\newblock \enquote{The deviant dynamics of death in heterogeneous populations.}
\newblock \emph{Sociological Methodology} 15:179--211.

\bibitem[\protect\citeauthoryear{Warren-Gash et~al.}{2011}]{warren-gash-11}
Warren-Gash, Charlotte, Krishnan Bhaskaran, Andrew Hayward, Gabriel~M. Leung,
  Su-Vui Lo, Chit-Ming Wong, Joanna Ellis, Richard Pebody, Liam Smeeth, and
  Benjamin~J. Cowling. 2011.
\newblock \enquote{Circulating influenza virus, climatic factors, and acute
  myocardial infarction: A time series study in {E}ngland and {W}ales and
  {H}ong {K}ong.}
\newblock \emph{Journal of Infectious Diseases} 203(12):1710--1718.

\bibitem[\protect\citeauthoryear{Warren-Gash et~al.}{2012}]{warren-gash-12}
Warren-Gash, Charlotte, Andrew~C. Hayward, Harry Hemingway, Spiros Denaxas,
  Sara~L. Thomas, Adam~D. Timmis, Heather Whitaker, and Liam Smeeth. 2012.
\newblock \enquote{Influenza infection and risk of acute myocardial infarction
  in {E}ngland and {W}ales: A {CALIBER} self-controlled case series study.}
\newblock \emph{Journal of Infectious Diseases} 206(11):1652--1659.

\bibitem[\protect\citeauthoryear{Warren-Gash et~al.}{2009}]{warren-gash-09}
Warren-Gash, Charlotte, Liam Smeeth, and Andrew~C. Hayward. 2009.
\newblock \enquote{Influenza as a trigger for acute myocardial infarction or
  death from cardiovascular disease: A systematic review.}
\newblock \emph{Lancet Infectious Diseases} 9(10):601--610.

\bibitem[\protect\citeauthoryear{Webster et~al.}{1992}]{webster-92}
Webster, Robert~G., William~J. Bean, Owen~T. Gorman, Thomas~M. Chambers, and
  Yoshihiro Kawaoka. 1992.
\newblock \enquote{Evolution and ecology of influenza {A} viruses.}
\newblock \emph{Microbiological Reviews} 56(1):152--179.

\bibitem[\protect\citeauthoryear{Wilkinson et~al.}{2004}]{wilkinson-04}
Wilkinson, Paul, Sam Pattenden, Ben Armstrong, Astrid Fletcher, R.~Sari Kovats,
  Punam Mangtani, and Anthony~J. McMichael. 2004.
\newblock \enquote{Vulnerability to winter mortality in elderly people in
  {B}ritain: Population based study.}
\newblock \emph{British Medical Journal} 329(7467):647.

\bibitem[\protect\citeauthoryear{Woods et~al.}{1989}]{woods-89}
Woods, R.~I., P.~A. Watterson, and J.~H. Woodward. 1989.
\newblock \enquote{The causes of rapid infant mortality decline in {E}ngland
  and {W}ales, 1861--1921. Part II.}
\newblock \emph{Population Studies} 43(1):113--132.

\bibitem[\protect\citeauthoryear{Yang et~al.}{2012}]{yang-12}
Yang, Jun, Chun-Quan Ou, Yan Ding, Ying-Xue Zhou, and Ping-Yan Chen. 2012.
\newblock \enquote{Daily temperature and mortality: A study of distributed lag
  non-linear effect and effect modification in {G}uangzhou.}
\newblock \emph{Environmental Health} 11(1):art.\@ no.\@ 63.

\bibitem[\protect\citeauthoryear{Yorke et~al.}{1979}]{yorke-79}
Yorke, James~A., Neal Nathanson, Giulio Pianigiani, and John Martin. 1979.
\newblock \enquote{Seasonality and the requirements for perpetuation and
  eradication of viruses in populations.}
\newblock \emph{American Journal of Epidemiology} 109(2):103--123.

\bibitem[\protect\citeauthoryear{Zhao et~al.}{2015}]{zhao-15}
Zhao, Zhongwei, Yuan Zhu, and Edward Jow-Ching Tu. 2015.
\newblock \enquote{Daily mortality changes in {T}aiwan in the 1970s: An
  examination of the relationship between temperature and mortality.}
\newblock \emph{Vienna Yearbook of Population Research} 13:71--90.

\end{thebibliography}
}
\end{multicols}


\special{"
[/Title (Summertime, and the livin' is easy)
/Author (Andrew Noymer noymer@uci.edu)
/Subject (demography)
/Keywords (mortality)
/Creator ()
/Producer()
/ModDate(2017)
/DOCINFO pdfmark
}

\end{document}